\documentclass[prd,nofootinbib,preprint,superscriptaddress]{revtex4}
\pdfoutput=1
\usepackage[T1]{fontenc}
\usepackage{amsmath,amssymb}
\usepackage{epsfig}
\usepackage{subfigure}
\usepackage{float}
\usepackage{graphicx}
\usepackage{bigints}
\usepackage[usenames,dvipsnames]{color}
\usepackage{slashed}
\usepackage{multirow}
\usepackage[colorlinks,citecolor=blue]{hyperref}
\usepackage{pdfpages}
\usepackage{color}
\usepackage{comment}
\begin{document}
\title{Successful cogenesis of baryon and dark matter from memory-burdened PBH}

\author{Debasish Borah}
\email{dborah@iitg.ac.in}
\affiliation{Department of Physics, Indian Institute of Technology Guwahati, Assam 781039, India}

\author{Nayan Das}
\email{nayan.das@iitg.ac.in}
\affiliation{Department of Physics, Indian Institute of Technology
Guwahati, Assam 781039, India}

\begin{abstract}
We study the possibility of producing the observed baryon asymmetry of the Universe (BAU) and dark matter (DM) from evaporating primordial black holes (PBH) beyond the semi-classical regime incorporating the impact of memory burden. In the simplest scenario of baryogenesis via vanilla leptogenesis with hierarchical right handed neutrino (RHN), it is possible to generate the observed BAU with memory-burdened PBH being sole contributor to the production of RHN. While it is not possible to achieve cogenesis in this minimal setup due to structure formation constraints on relic allowed DM parameter space, we show the viability of successful cogenesis in the resonant leptogenesis regime. We also show that successful cogenesis can be achieved in a simple baryogenesis model without taking the leptogenesis route. Due to the possibility of generating asymmetry even below the sphaleron decoupling era, the direct baryogenesis route opens up new parameter space of memory-burdened PBH. The two scenarios of successful cogenesis can also be distinguished by observations of stochastic gravitational waves produced from PBH density fluctuations.
\end{abstract}
\maketitle
\section{Introduction}
\label{sec:Intro}
Primordial black holes (PBH), originally proposed by Zeldovich \cite{Zeldovich:1967lct} and later by Hawking \cite{Hawking:1974rv, Hawking:1975vcx} can be very interesting from cosmological perspectives \cite{Chapline:1975ojl, Carr:1976zz}. A comprehensive recent review of PBH can be found in \cite{Carr:2020gox}. While PBH parameters namely, its initial mass $(m_{\rm in})$ and initial energy fraction $(\beta)$ are tightly constrained from cosmology and astrophysics related bounds, the ultra-light PBH mass window is relatively less constrained as such PBH evaporate by emitting Hawking radiation \cite{Hawking:1974rv, Hawking:1975vcx} before the big bang nucleosynthesis (BBN) epoch. Quantitatively, this ultra-light mass window corresponds to $0.5\,\text{g}\lesssim m_\text{in}\lesssim 3.4\times 10^8\,\text{g}$ where the lower bound arises due to the cosmic microwave background (CMB) limits on the scale of inflation. However, if such PBH do not dominate the energy density of the Universe, implying a small value of $\beta$, then the upper bound does not apply. Evaporation of such ultra-light PBH can lead to production of baryon asymmetry of the Universe (BAU) \cite{Hawking:1974rv, Carr:1976zz, Baumann:2007yr, Hook:2014mla, Fujita:2014hha, Hamada:2016jnq, Morrison:2018xla, Hooper:2020otu, Perez-Gonzalez:2020vnz, Datta:2020bht, JyotiDas:2021shi, Smyth:2021lkn, Barman:2021ost, Bernal:2022pue, Ambrosone:2021lsx, Calabrese:2023key, Ghoshal:2023fno, Barman:2024slw} and dark matter (DM) \cite{Morrison:2018xla, Gondolo:2020uqv, Bernal:2020bjf, Green:1999yh, Khlopov:2004tn, Dai:2009hx, Allahverdi:2017sks, Lennon:2017tqq, Hooper:2019gtx, Chaudhuri:2020wjo, Masina:2020xhk, Baldes:2020nuv, Bernal:2020ili, Bernal:2020kse, Lacki:2010zf, Boucenna:2017ghj, Adamek:2019gns, Carr:2020mqm, Masina:2021zpu, Bernal:2021bbv, Bernal:2021yyb, Samanta:2021mdm, Sandick:2021gew, Cheek:2021cfe, Cheek:2021odj, Friedlander:2023qmc, Borah:2024lml}. The common origin of BAU via leptogenesis \cite{Fukugita:1986hr} and DM from evaporating PBH has been studied in several works \cite{Fujita:2014hha, Datta:2020bht, JyotiDas:2021shi,Barman:2021ost, Barman:2022gjo}. Baryogenesis scenarios without the leptogenesis route were also discussed in the context of evaporating PBH by the authors of \cite{Barman:2022pdo, Choi:2023kxo}. While the requirement of successful leptogenesis corners the PBH mass to $\sim\mathcal{O}(1)$ g, in a direct baryogenesis setup we can have PBH mass as large as $\sim\mathcal{O}(10^7)$ g due to the possibility of producing BAU directly below sphaleron decoupling temperature. It was also shown that the resulting stochastic gravitational waves (GW) from PBH density fluctuations peak around mHz-kHz frequencies due to larger allowed masses of PBH in the baryogenesis setup \cite{Barman:2022pdo}. This keeps the setup verifiable at GW detectors and distinguishable from the leptogenesis scenarios.

The PBH originated cogenesis predictions obtained in the works mentioned above can however, change significantly if we deviate from the assumption that the PBH evaporation continues to obey the semi-classical approximation of Hawking \cite{Hawking:1974rv, Hawking:1975vcx} till the end of its lifetime. In fact, Hawking's semi-classical approximations ignore the backreaction of the emitted particles, which can be significant once the energy of the
emitted quanta becomes comparable to the PBH energy. Recently, it was pointed out that such backreactions lead to a \textit{memory burden} effect which slows down the evaporation \cite{Dvali:2018xpy, Dvali:2018ytn, Dvali:2024hsb}. This effect becomes more prominent as PBH mass falls below a certain threshold, leading to an enhanced lifetime. This has interesting phenomenological consequences for dark matter, gravitational waves, cogenesis as well as high energy astroparticle physics \cite{Dvali:2021byy, Balaji:2024hpu, Barman:2024iht, Bhaumik:2024qzd, Barman:2024ufm, Kohri:2024qpd, Jiang:2024aju, Zantedeschi:2024ram, Chianese:2024rsn, Barker:2024mpz, Alexandre:2024nuo, Thoss:2024hsr, Haque:2024eyh}. In cogenesis related work \cite{Barman:2024iht}, it was shown that memory-burdened PBH can not simultaneously lead to DM and BAU via vanilla leptogenesis with hierarchical right handed neutrino (RHN). While correct DM abundance requires PBH mass $\geq \mathcal{O}(10^3 \, \rm g)$, correct BAU via vanilla leptogenesis can be generated for PBH mass $\leq \mathcal{O}(10^3 \, \rm g)$. The primary hurdle in achieving both simultaneously arises from structure formation constraints on relic allowed DM parameter space. In this work, we propose two alternatives to alleviate it namely via resonant leptogenesis and direct baryogenesis without the leptogenesis route. We show the viability of cogenesis in these two scenarios while keeping PBH parameters in a range allowed from astrophysics and cosmological bounds. We also show that these two scenarios of succeesful cogenesis from memory-burdened PBH can be distinguished via observations of stochastic GW backgrounds. While leptogenesis requires generation of asymmetry before the sphaleron decoupling epoch $T_{\rm sph} \sim 130$ GeV, direct baryogenesis can occur at temperatures as low as a few tens of MeV. This alters the allowed PBH paramater space affecting the predictions for GW produced from PBH density fluctuations. 

This paper is organised as follows. In section \ref{sec1}, we summarise PBH evaporation in semi-classical regime and with memory-burden effect. In section \ref{sec2}, we discuss the results of cogenesis followed by discussion of gravitational waves detection prospects in section \ref{sec3}. We finally conclude in section \ref{sec4}.

\section{PBH evaporation}
\label{sec1}
We assume the formation of PBH in a radiation dominated Universe after inflation. In the early universe, PBH can be formed in a variety of ways like, from inflationary perturbations \cite{Hawking:1971ei, Carr:1974nx, Wang:2019kaf, Byrnes:2021jka, Braglia:2022phb}, first-order phase transition (FOPT) \cite{Crawford:1982yz, Hawking:1982ga, Moss:1994iq, Kodama:1982sf}, the collapse of topological defects \cite{Hawking:1987bn, Deng:2016vzb} etc. We remain agnostic about such formation mechanisms and consider PBH to have a monochromatic mass function with initial mass $M_{\rm in}$ and energy fraction
\begin{eqnarray}
    \beta \equiv \frac{\rho_{\rm BH} (T_{\rm in})}{\rho_{R}(T_{\rm in})},
\end{eqnarray}
where $T_{\rm in}$ is the temperature during PBH formation, and $\rho_{\rm BH}$, $\rho_{\rm R}$ are the PBH and radiation energy density respectively. We also consider the PBH to be of Schwarzschild type having no spin and charge. Considering PBH formation in radiation dominated universe, the mass of PBH from gravitational collapse is
typically close to the mass enclosed by the post-inflationary particle horizon and is given by
\begin{equation}
     M_{\rm in}=\gamma \frac{4\,\pi\,}{3\,\mathcal{H}\left(T_\text{in}\right)^{3}}\,\rho_\text{R}\left(T_\text{in}\right)\, ,
\end{equation}
where $\mathcal{H}$ represents the Hubble expansion rate. The factor $\gamma$ is associated with the uncertainty of PBH formation with typical value of $\simeq 0.2$ \cite{Carr:1974nx}.  Given that PBH forms during early radiation dominated era, the epoch of formation can be written as 
\begin{eqnarray}
    t_{\rm in} = \frac{M_{\rm in}}{8\, \pi \gamma M_{P}^2},
\end{eqnarray}
with $M_{P}$ denoting the reduced Planck mass. The corresponding SM plasma temperature $T=T_\text{in}$ is given by
\begin{equation}
T_\text{in}=\Biggl(\frac{1440\,\gamma^2}{g_\star\left(T_\text{in}\right)}\Biggr)^{1/4}\,\sqrt{\frac{M_P}{M_\text{in}}}\,M_P\,.
\label{eq:pbh-in}
\end{equation}
The instantaneous temperature and entropy associated with a black hole of mass $M_{\rm BH}$ are given as 
\begin{eqnarray}
    T_{\rm BH} &=& \frac{M_{P}^2}{M_{\rm BH}}, \,\,\,\,
    S = \frac{1}{2} \left(\frac{M_{\rm BH}}{M_{P}}\right)^2 = \frac{1}{2} \left(\frac{M_{P}}{T_{\rm BH}}\right)^2, 
    \label{eq:entropy}
\end{eqnarray}
respectively. After formation, PBH can evaporate by emitting Hawking radiation \cite{Hawking:1974rv, Hawking:1975vcx}. We summarise the details of PBH evaporation using Hawking's semi-classical approximation and memory-burden effect as follows.

\textbf{Semi-classical (SC) regime: } In the semi-classical approximation, the mass loss rate is given by \cite{MacGibbon:1991tj}
\begin{eqnarray}
    \frac{dM_{\rm BH}}{dt} = - \epsilon \frac{M^4_{P}}{M^2_{\rm BH}},
    \label{eq:massloss}
\end{eqnarray}
where 
\begin{equation}
\epsilon = \frac{27}{4} \frac{\pi g_{*, H}(T_{\rm BH})}{480}, \,\, g_{*, H}(T_{\rm BH}) = \sum_i \omega_i g_{i,H}, \,\,g_{i,H}=
    \begin{cases}
        1.82
        &\text{for }s=0\,,\\
        1.0
        &\text{for }s=1/2\,,\\
        0.41
        &\text{for }s=1\,,\\
        0.05
        &\text{for }s=2\,,\\
    \end{cases}
    \label{eqn:gsh}
\end{equation}
with $\omega_i=2s_i+1$ for massive particles of spin $s_i$, $\omega_i=2$ for massless species with $s_i>0$, and $\omega_i=1$ for spinless species. Integrating Eq. \eqref{eq:massloss}, PBH mass at an epoch $t > t_{\rm in}$ can be obtained as 
\begin{eqnarray} \label{eq:MBH_time_SC}
    M_{\rm BH}(t) = M_{\rm in} \left(1-\frac{3\,\epsilon\,M_{P}^4}{M_{\rm in}^3}(t-t_{\rm in})\right)^{1/3} \equiv  M_{\rm in} \left(1- \Gamma^0_{\rm BH}(t-t_{\rm in})\right)^{1/3}.
\end{eqnarray}
Here, $\Gamma^0_{\rm BH}$ can be termed as the associated decay width for the semi-classical regime. If the semi-classical regime is valid till the complete evaporation of PBH, its lifetime $t^0_{\rm ev}\gg t_{\rm in}$ is given as
\begin{eqnarray} \label{eq:t0ev}
    t^0_{\rm ev} = \frac{1}{\Gamma^0_{\rm BH}} = \frac{M_{\rm in}^3}{3\,\epsilon\,M_{P}^4}.
\end{eqnarray}
The corresponding evaporation temperature can then be computed taking into account $\mathcal{H}(T_\text{evap})\sim\frac{1}{(t^0_{\rm ev})^2}\sim\rho_\text{R}(T_\text{evap})$ as
\begin{equation}
T_\text{evap}\equiv\Bigl(\frac{90\,M_P^2}{4\,\pi^2\,g_\star\left(T_\text{evap}\right)\, (t^0_{\rm ev})^2}\Bigr)^{1/4}\,.
\label{eq:pbh-Tev}
\end{equation}
However, if the PBH component dominates the total energy density of the universe at some epoch, the SM temperature just after the complete evaporation of PBHs is: $\overline{T}_\text{evap}=2/\sqrt{3}\,T_\text{evap}$~\cite{Bernal:2020bjf}.

\textbf{Memory-burdened (MB) regime:} After formation, PBH starts radiating particles in a self-similar semi-classical process until it loses certain fraction of its original mass. We quantify the end of semi-classical regime via the condition $M_{\rm BH}=q\,M_{\rm in}$, where $0<q<1$. Once PBH mass reduces to $q M_{\rm in}$, quantum memory effects starts to dominate altering PBH evaporation rate. Thus, the evaporation takes place in two different regimes namely, i) semi-classical regime for PBH mass range $(M_{\rm in}, qM_{\rm in})$ and ii) memory-burdened regime for PBH mass range $(q M_{\rm in}, 0)$.
The PBH evaporation rate in the second regime is given as
\begin{eqnarray}
     \frac{dM_{\rm BH}}{dt} = - \frac{\epsilon}{[S(M_{\rm BH})]^k} \frac{M^4_{P}}{M^2_{\rm BH}},
\end{eqnarray}
where $S(M_{\rm BH})$ is same as the one defined in Eq. \eqref{eq:entropy}. Integrating this from an initial mass $q M_{\rm in}$ at $t=t_{q}$ to a later time $t$, we get
\begin{eqnarray} \label{eq:MBH_time_MB}
    M_{\rm BH} (t) = q M_{\rm in} \left(1- \Gamma^k_{\rm BH}(t-t_{q})\right)^{1/(3+2k)},
\end{eqnarray}
where 
\begin{eqnarray}
    \Gamma^{k}_{\rm BH} \equiv 2^k (3+2k) \, \epsilon \, M_{\rm P} \left(\frac{M_{P}}{qM_{\rm in}}\right)^{3+2k}
\end{eqnarray}
is the associated decay width for the MB regime. The total lifetime of a PBH then can be written as 
\begin{eqnarray}
    t^k_{\rm ev} = t_{q} + \frac{1}{\Gamma^{k}_{\rm BH}} = \frac{1-q^3}{\Gamma^0_{\rm BH}}  + \frac{1}{\Gamma^{k}_{\rm BH}}. 
\end{eqnarray}
Due to the memory-burden effect, the SC regime is valid until $M_{\rm BH} = q M_{\rm in}$. The corresponding time $t_{q}$ can be obtained as (assuming $t_{q} \gg t_{\rm in}$)
\begin{eqnarray}
    t_{q} = \frac{1-q^3}{\Gamma^0_{\rm BH}},
\end{eqnarray}
which reduces to Eq. \eqref{eq:t0ev} for $q \to 0$. 

\begin{figure}
    \centering
    \includegraphics[width=0.48\linewidth]{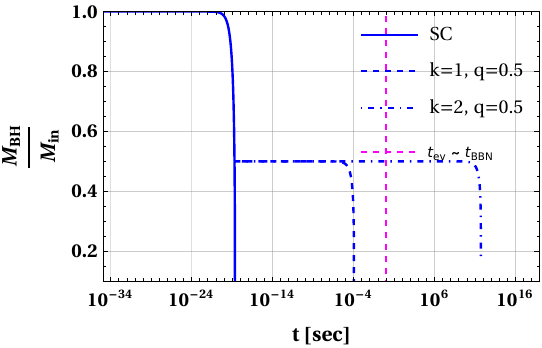}
    \includegraphics[width=0.48\linewidth]{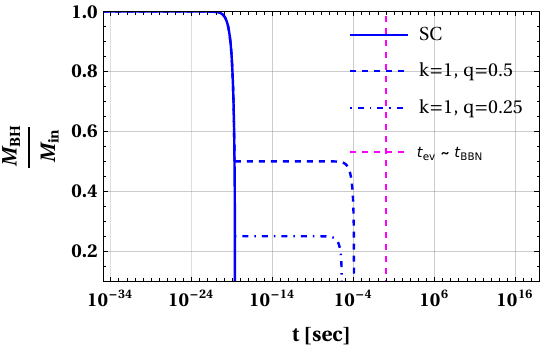}
    \caption{Evolution of PBH mass fraction for two different values of $k$ with $q=0.5$ in the left panels and for two different values of $q$ with $k=1$ in the right panel. The PBH mass for both the panel is taken to be $10^{3}$ g. The vertical dashed line represents the time corresponding to the BBN epoch.}
    \label{fig1}
\end{figure}

The critical value of $\beta$, denoted as $\beta_{\rm c}$ can be obtained as 
\begin{eqnarray}
    \beta_{\rm c} = \left(\frac{(3+2k)2^{k}\epsilon}{8\, q^3 \pi \gamma}\right)^{1/2} \left(\frac{M_{P}}{q M_{\rm in}}\right)^{1+k}.
\end{eqnarray}
For $\beta > \beta_{c}$, PBH dominates the energy density of the Universe before complete evaporation, leading to an early matter dominated era. Given PBH domination before evaporation, the Hubble expansion rate during evaporation can be written as 
\begin{eqnarray}
    \mathcal{H} (a_{\rm ev}) = \frac{2}{3}\frac{1}{t^k_{\rm ev}} = \sqrt{\frac{\rho_{\rm R}(a_{\rm ev})}{3 M_{P}^2}},
\end{eqnarray}
which give us the evaporation temperature as 
\begin{equation}
    T_{\rm ev}= M_{P} \left(\frac{4}{3\alpha}\right)^{1/4} \left(\frac{3\times2^{k}(3+2k)\epsilon \left(\frac{M_{P}}{M_{\rm in}}\right)^{3+2k}}{3\times q^{3+2k}+(1-q^3)2^{k}(3+2k)\left(\frac{M_{P}}{M_{\rm in}}\right)^{2k}}\right)^{1/2}.
\end{equation}
Here $\alpha = \frac{\pi^2}{30} \, g_{*} (T_{\rm ev})$, with $g_{*} (T_{\rm ev})$ being the relativistic degrees of freedom associated with SM bath at $T=T_{\rm ev}$. For either $q\to0$ or $k\to 0$, the above expression reduces to standard semi-classical expression. In this work, we focus on the scenario $\beta>\beta_{c}$ i.e. PBH dominates the energy density of the Universe for a finite epoch before complete evaporation. Since PBH energy density gets converted into radiation density during evaporation, we use the relation $\rho_{\rm BH}(a_{\rm ev}) = \rho_{\rm R} (a_{\rm ev}) \simeq n_{\rm BH}(a_{\rm ev}) \times q M_{\rm in}$ resulting in the number density of PBH at the evaporation temperature as
\begin{equation}  \label{eq:nBH}
    n_{\rm BH} (a_{\rm ev}) = \frac{4}{3} M^4_{P} \left(\frac{3\times2^{k}(3+2k)\epsilon \left(\frac{M_{P}}{M_{\rm in}}\right)^{3+2k}}{3\times q^{3+2k}+(1-q^3)2^{k}(3+2k)\left(\frac{M_{P}}{M_{\rm in}}\right)^{2k}}\right)^{2} \frac{1}{q M_{\rm in}}.
\end{equation}
The slight difference in the expressions for $T_{\rm ev}, n_{\rm BH}(a_{\rm ev})$ compared to \cite{Barman:2024iht} is due to the fact that we are not using any approximations. However, such differences do not lead to any significant changes in numerical calculations which we discuss below. Fig. \ref{fig1} shows the differences in the semi-classical and memory-burdened regimes in terms of PBH mass-loss rate. Clearly, with non-zero values of $k, q$, the evaporation rate slows down towards the later epochs while coinciding with the semi-classical predictions initially.

\section{Cogenesis of dark and visible matter}
\label{sec2}
Number of particles with mass $m_{j}$ (for species $j$ with internal degrees of freedom $g_j$) produced by complete evaporation of a PBH of initial mass $M_{\rm in}$ can be derived using memory burden effect as \cite{Haque:2024eyh,Barman:2024iht}
\begin{equation} \label{eq:Nj}
    N_{j} \simeq \frac{27}{128} \frac{\xi g_{j} \zeta (3)}{\pi^3 \epsilon} 
    \begin{cases}
    \left(\frac{M_{\rm in}}{M_P}\right)^{2},& \text{if $m_{j} < T^{\rm in}_{\rm BH}$} \\
    \left(\frac{M_P}{m_{j}}\right)^{2},& \text{if $m_{j} > T^{\rm in}_{\rm BH}$}.
\end{cases}
\end{equation}
Here, $\xi$ takes value of $1$ and $3/4$ for bosons and fermions respectively. Note that the above expression is independent of both $k$ and $q$ as we are considering complete evaporation of PBH. See appendix \ref{appen1} for the details of particle production from memory-burdened PBH. Two different cases arise depending upon the relative size of mass $m_{j}$ and initial Hawking temperature $T^{\rm in}_{\rm BH}$ as species satisfying $m_{j} < T^{\rm in}_{\rm BH}$ can be produced from PBH evaporation from the very beginning while species with $m_{j} > T^{\rm in}_{\rm BH}$ can be produced only after instantaneous Hawking temperature increases beyond $m_j$.

Without any loss of generality, we consider a real scalar singlet DM having no coupling to the SM bath particles. Therefore, it is produced only gravitationally from PBH evaporation. For a single PBH producing $N_{\rm DM}$ number of DM species during the entire period of evaporation, the total DM relic can be estimated as
\begin{equation}
    \Omega_{\rm DM}h^2 = 1.6\times 10^{8} \frac{g_{*s}(T_{0})}{g_{*s}(T_{\rm ev})} \frac{n_{\rm BH}(T_{\rm ev})}{T^3_{\rm ev}} \frac{m_{\rm DM}}{\rm GeV} N_{\rm DM}.
\end{equation}

\begin{figure}[htb]
    \centering
    \includegraphics[width=0.48\linewidth]{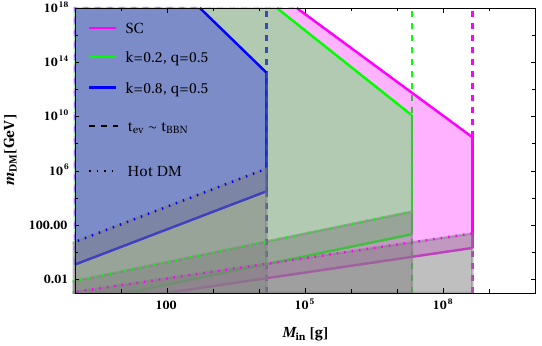}
    \includegraphics[width=0.48\linewidth]{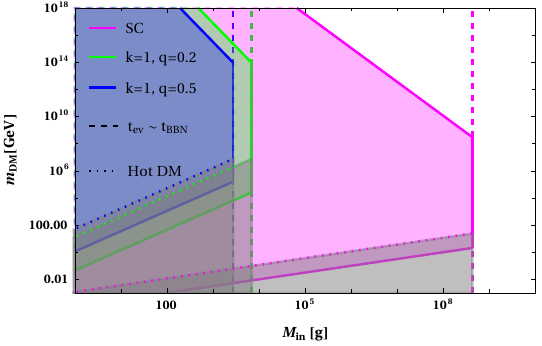}
    \caption{$m_{\rm DM}$ as a function of $M_{\rm in}$ for different values of $q$ in the left panel and different values of $k$ in the right panel. In the shaded region, the DM abundance is overproduced. Correct DM abundance is produced at the boundary of the shaded region. The vertical dashed lines corresponds to $t_{\rm ev} = t_{\rm BBN}$. The grey shaded region is disfavoured by structure formation constraints.}
    \label{fig:DM}
\end{figure}

Fig. \ref{fig:DM} shows possible DM mass as a function of initial PBH mass for different values of \{$k, q$\}. The shaded regions for each benchmark correspond to DM overproduction. The region towards the right of vertical dashed line are disallowed from BBN requirement $t_{\rm ev} < t_{\rm BBN}$. In the later part of our work, we primarily focus on the case $m_{\rm DM} > T^{\rm in}_{\rm BH}$. This helps in evading the structure formation constraint unlike the case $m_{\rm DM} < T^{\rm in}_{\rm BH}$. The grey shaded regions in Fig. \ref{fig:DM} are ruled out from structure formation constraints on DM free-streaming length\cite{Diamanti:2017xfo, Fujita:2014hha,Masina:2020xhk,Barman:2022gjo}. This essentially rules out the low mass regime of the relic allowed DM parameter space for each of the benchmark values of $\{k, q\}$. The details related to the structure formation constraints can be found in appendix \ref{appen2}.

In addition to the production of DM from PBH evaporation, DM is also produced from gravity mediated process. The details of the calculations are provided in Appendix \ref{appen3}. As shown in Fig. \ref{fig:GM_DM} of Appendix \ref{appen3}, the parameter space where such gravity mediated process satisfies correct DM relic already leads to DM overproduction solely from PBH evaporation. In other words, contribution from such gravity mediated process is sub-dominant in the region of parameter space where PBH evaporation produces correct DM relic. This justifies the validity of our numerical results considering production of particles only from PBH evaporation. However, this conclusion is valid only for reheat temperature $T_{\rm RH}=T_{\rm in}$, used in the estimate of particle production from gravity mediated process.

\subsection{Leptogenesis from PBH}
\begin{figure}
    \centering
    \includegraphics[width=0.48\linewidth]{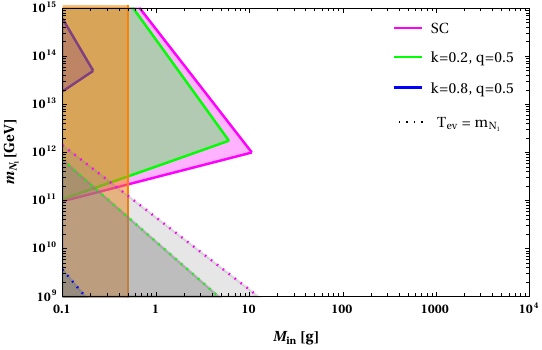}
    \includegraphics[width=0.48\linewidth]{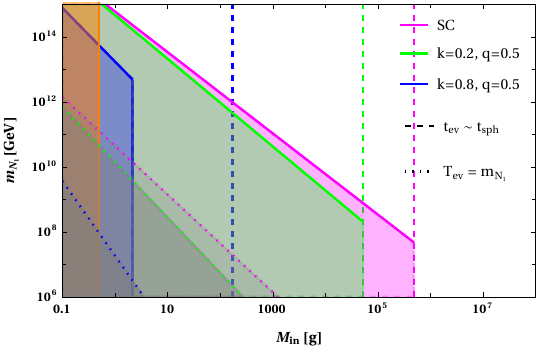}
    \caption{Leptogenesis from PBH: hierarchical (left panel) and resonant (right panel). In the shaded regions, the observed baryon asymmetry can be produced. The vertical dashed lines (right panel) correspond to $t_{\rm ev} = t_{\rm sph}$. The dotted lines denote $T_{\rm ev} = m_{N_{1}}$.  PBH mass $m_{\rm in} \lesssim 0.5$ g is inconsistent with CMB data and is shown by the shaded orange region.}
    \label{fig:Lep}
\end{figure}

Leptogenesis \cite{Fukugita:1986hr} is a popular way to generate the observed baryon asymmetry. Instead of creating baryon asymmetry directly, a non-zero lepton asymmetry is first created which gets converted into baryon asymmetry via $(B+L)$-violating electroweak sphaleron transitions~\cite{Kuzmin:1985mm}. Interestingly, the minimal leptogenesis frameworks also explains the origin of light neutrino mass via seesaw mechanism. The simplest of these scenarios is the Type-I seesaw mechanism \cite{Mohapatra:1979ia,Yanagida:1979as,GellMann:1980vs,Glashow:1979nm} where we have three gauge singlet right handed neutrinos. The relevant Yukawa Lagrangian can be written as
\begin{equation}
    -\mathcal{L} \supset y_{\alpha i} \overline{L_\alpha} \tilde{H} N_i + \frac{1}{2} M_{N_i} \overline{N^c_i}N_i + {\rm h.c.}
\end{equation}
Considering hierarchical RHN mass spectrum $M_{N_1} \ll M_{N_2} \ll M_{N_3}$, it is essentially the CP violating out-of-equilibrium decay of the lightest RHN $N_1$, which generates the lepton asymmetry. For hierarchical RHNs, the  CP asymmetry parameter has an upper bound \cite{Davidson:2002qv}
\begin{eqnarray}
    \epsilon^{\Delta L}_{1} = \frac{3 M_{N_{1}}\sqrt{(\Delta m_{\rm atm})^2}}{4 \pi v^2}.
\end{eqnarray}
Here, $v/\sqrt{2}=174$ GeV is the vacuum expectation value (VEV) of the SM Higgs $H$ and $(\Delta m_{\rm atm})^2 \simeq 2.4\times 10^{-3} \rm eV^{2}$ is the atmospheric mass-squared difference of light neutrinos. Considering non-thermal production of RHN solely from PBH evaporation, the final baryon asymmetry can be estimated as \cite{Fujita:2014hha,Datta:2020bht,Barman:2021ost,Barman:2022gjo}
\begin{eqnarray}
    Y^{\Delta L}_{\rm B} (T_{0})=\frac{n_B}{s}\Big|_{T_0} = \epsilon^{\Delta L}_{1}\, a_{\rm sph} \frac{n_{\rm BH}(a_{\rm ev}) }{s (a_{\rm ev})} N_{N_{1}}, 
\end{eqnarray}
where $N_{N_{1}}$ denotes the number of $N_{1}$  produced from the complete evaporation of one PBH and $a_{\rm sph}=28/79$ represents the sphaleron conversion factor. The left panel of Fig. \ref{fig:Lep} shows the parameter space in $M_{N_1}-M_{\rm in}$ plane consistent with successful leptogenesis with hierarchical RHNs. The shaded regions towards the left of solid lines for different benchmark choices of $k, q$ are consistent with successful leptogenesis. The bottom left triangular regions correspond to $M_{N_1} < T_{\rm ev}$ inconsistent with non-thermal leptogenesis requirement. Clearly, the allowed parameter space shrinks after inclusion of the memory burden effect.

The lepton asymmetry can be significantly enhanced if two of the RHNs are quasi-degenerate, i.e., $\Delta M=M_{N_2}-M_{N_1} \ll \overline{M}=(M_{N_1}+M_{N_2})/2$ leading to the resonant leptogenesis scenario \cite{Pilaftsis:2003gt, Dev:2017wwc}. For $\Delta M \simeq \Gamma$, the CP asymmetry parameter can be as large as $\mathcal{O}(1)$, where $\Gamma$ indicates the decay width of the RHN. Due to resonantly enhanced CP asymmetry, the allowed parameter space broadens up as shown in the right panel of Fig. \ref{fig:Lep}. The description of the shaded regions remains same as the left panel. The vertical dashed lines correspond to $t_{\rm ev} = t_{\rm sph}$, the sphaleron decoupling epoch. Therefore, the region towards the right of these dashed lines are disallowed for respective benchmark choice of $k, q$ as the asymmetry is generated below sphaleron decoupling and can not be converted into baryon asymmetry.

\subsection{Baryogenesis from PBH}
Instead of producing a non-zero lepton asymmetry first, it is also possible to generate baryon asymmetry directly. The advantage of such a scenario is that it no longer requires $t_{\rm ev} < t_{\rm sph}$ as long as PBH evaporation takes place before the BBN. To realise baryogenesis, we follow a simple particle physics setup similar to the earlier works~\cite{Allahverdi:2010im, Allahverdi:2010rh,Allahverdi:2013tca, Allahverdi:2013mza,Allahverdi:2017edd}. Two copies of color triplet, $SU(2)$ singlet scalars $S_{1,2}$ of hypercharge $2/3$ are introduced whose decay generates the baryon asymmetry. A gauge singlet chiral fermion $\psi$ is also included to ensure non-zero CP asymmetry. The relevant part of the Yukawa Lagrangian can be written as \cite{Allahverdi:2013mza, Allahverdi:2017edd, Dev:2015uca, Davoudiasl:2015jja}
\begin{align}\label{eq:lg}
& -\mathcal{L}\supset \lambda\,S\,\psi\,u^c + \lambda'\,S^\star\,d^c\,d^c + \frac{1}{2}\, m_\psi\,\overline{\psi^c}\,\psi+\text{h.c.}\,,
\end{align}
where we have suppressed all the flavour indices. Clearly, the Majorana mass term of $\psi$ is the source of baryon number violation ($\Delta B=2$) in this model. Due to the presence of two different vertices of $S_{i}$ shown in Eq. \eqref{eq:lg}, non-zero CP asymmetry $\epsilon_i$ arises. Keeping $S_i$, in the non-thermal ballpark which keeps wash-out and dilution processes sub-dominant, the net baryon asymmetry at present epoch can be estimated as
\begin{equation}
   Y_B(T_0) \equiv Y_B(T_{\rm ev}) = \left(\epsilon_1+\epsilon_2\right)\,\frac{n_S}{s}\Bigg|_{T_{\rm ev}} = (\epsilon_{1} + \epsilon_{2}) N_{S} \frac{n_{\rm BH}(T_{\rm ev})}{s(T_{\rm ev})}. 
\end{equation}
Here we consider $m_{S_1} \approx m_{S_2} \equiv m_S$ leading to $N_{S_1}=N_{S_2}=N_S$, the number of $S_{1,2}$ produced from complete evaporation of a single PBH.

\begin{figure}
    \centering
    \includegraphics[width=0.48\linewidth]{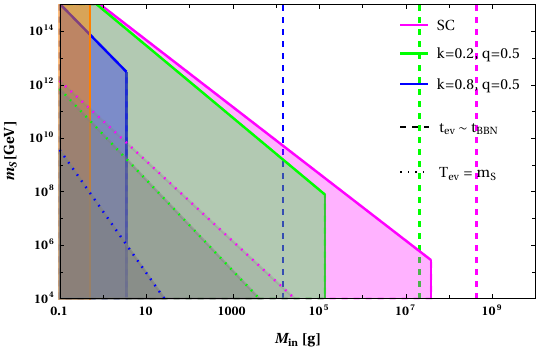}
    \includegraphics[width=0.48\linewidth]{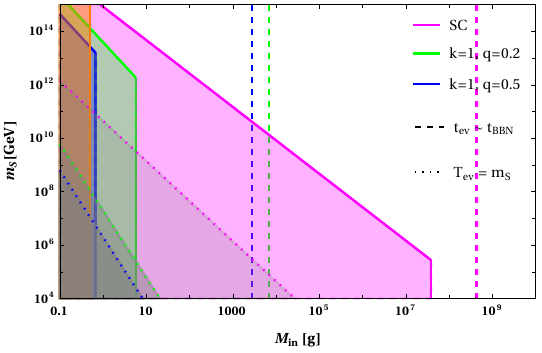}
    \caption{Baryogenesis from PBH: $m_{S}$ as a function of $M_{\rm in}$ for different value of $q$ in the left panel and different values of $k$ in the right panel. In the shaded region towards the left of solid lines, correct baryon asymmetry can be generated. The vertical dashed lines correspond to $t_{\rm ev} = t_{\rm BBN}$. The dotted lines denote $T_{\rm ev} = m_{S}$. PBH mass $m_{\rm in} \lesssim 0.5$ g is inconsistent with CMB data and is shown by the shaded orange region.}
    \label{fig:Baryo}
\end{figure}

Fig. \ref{fig:Baryo} shows the parameter space in $m_S-M_{\rm in}$ plane for baryogenesis, considering large CP asymmetry $\epsilon_i =0.1$. The shaded regions towards the left of the solid lines are consistent with the observed baryon asymmetry. The shaded regions in the bottom left triangular portion are inconsistent with non-thermal baryogenesis solely from PBH evaporation. The dashed vertical lines correspond to $t_{\rm ev} = t_{\rm BBN}$ such that the region towards the right is disfavoured. Comparing with the results of hierarchical as well as resonant leptogenesis shown in Fig. \ref{fig:Lep}, one can clearly see the enlarged parameter space in the baryogenesis setup allowing much larger initial PBH masses. This has crucial implications for the predictions of stochastic gravitational waves as we discuss in the upcoming section.

\subsection{Cogenesis}
\begin{figure}
    \centering
    \includegraphics[width=0.48\linewidth]{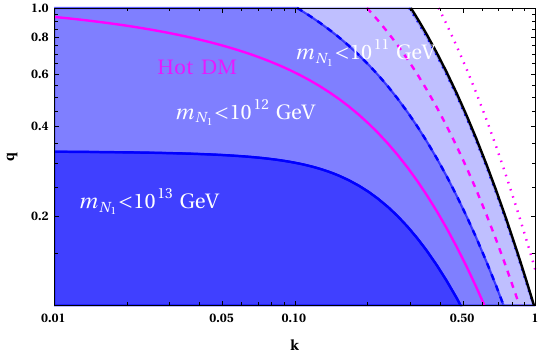}
     \includegraphics[width=0.48\linewidth]{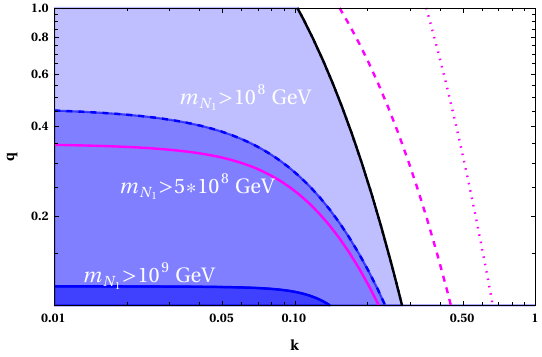}
    \caption{Left panel: Hierarchical leptogenesis with $M_{\rm in}=1$ g. Shaded regions are consistent with observed baryon asymmetry. The pink solid, dashed and dotted lines represent DM mass of 0.1 MeV, 1 MeV and 10 MeV respectively keeping DM hot, disfavoured from structure formation constraints. Right panel: Same as left panel but for resonant leptogenesis, $M_{\rm in}=10^5$ g. The pink solid, dashed and dotted lines represent relic allowed DM mass of $10^{18}$, $10^{16}$ and $10^{12}$ GeV respectively.}
    \label{fig:cogenesis1}
\end{figure}

Based on the above discussion of dark matter, leptogenesis and baryogenesis from PBH, we can summarise the viability of producing both dark and visible matter from evaporating PBH beyond the semi-classical approximations. Fig. \ref{fig:cogenesis1} summarises the results for leptogenesis in $(q,k)$ plane. The left panel shows the parameter space for hierarchical leptogenesis with $M_{\rm in}=1$ g. The shaded regions are consistent with the observed baryon asymmetry. The pink solid, dashed and dotted lines correspond to DM mass of 0.1 MeV, 1 MeV and 10 MeV respectively with correct relic. Such light DM produced from PBH evaporation are ruled out from structure formation constraints due to large free-streaming length. We can have relic allowed parameter space in heavy DM limit too, but it becomes super-Planckian and hence discarded. This concludes that cogenesis of DM and baryon asymmetry via hierarchical leptogenesis is not possible with memory-burdened PBH, confirming earlier results \cite{Barman:2024iht}. The right panel of the same figure shows the corresponding parameter space for resonant leptogenesis with  with $M_{\rm in}=10^5$ g. The pink solid, dashed and dotted lines represent DM mass of $10^{18}$, $10^{16}$ and $10^{12}$ GeV respectively. Clearly, such heavy DM remains safe from structure formation constraints. Therefore, cogenesis is viable in resonant leptogenesis scenario for a wide range of $(q,k)$ keeping PBH in memory-burdened regime.

\begin{figure}
    \centering
    \includegraphics[width=0.48\linewidth]{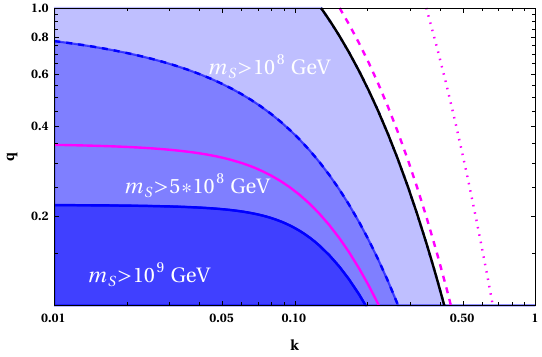}
    \includegraphics[width=0.48\linewidth]{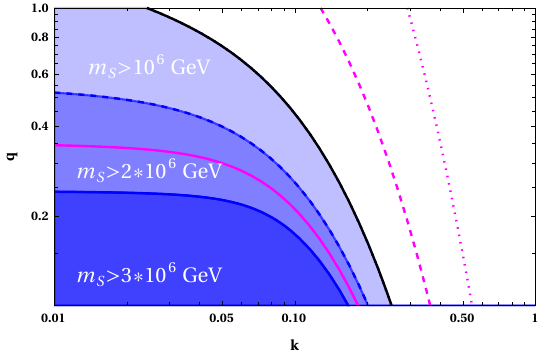}
    \caption{Left panel: Baryogenesis with $M_{\rm in} = 10^{5}$ g with the shaded regions being consistent with observed baryon asymmetry. The pink solid, dashed and dotted lines represents DM mass of $10^{18}$, $10^{16}$ and $10^{12}$ GeV respectively. Right panel: Same as left panel but with $M_{\rm in} = 10^{7}$ g. The pink solid, dashed and dotted lines represents DM mass of $10^{13}$, $10^{11}$ and $10^{9}$ GeV respectively.}
    \label{fig:cogenesis3}
\end{figure}

Fig. \ref{fig:cogenesis3} shows the parameter space in $(q,k)$ plane consistent with DM and baryogenesis. The shaded regions are consistent with successful baryogenesis while the pink colored lines correspond to relic allowed parameter space for DM with specific masses. Similar to resonant leptogenesis, here also DM remains superheavy and hence trivially satisfy structure formation bounds. 

\begin{figure}
    \centering
    \includegraphics[width=0.48\linewidth]{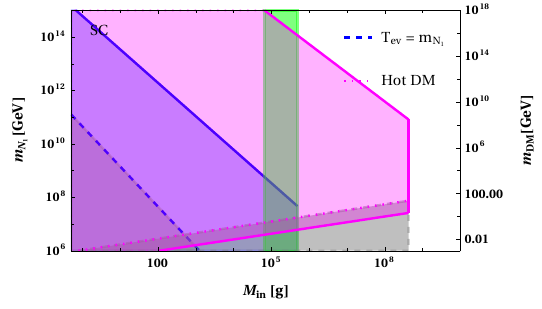}
    \includegraphics[width=0.48\linewidth]{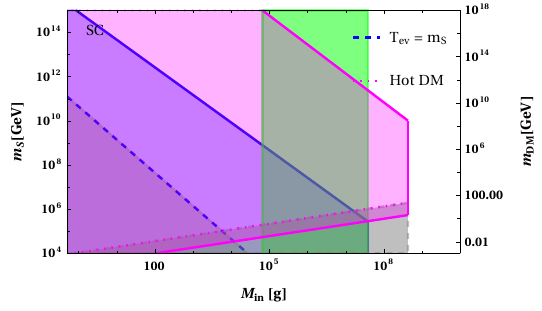}
    \caption{Left panel : $M_{\rm in}$ vs $m_{N_{1}}$ ($m_{\rm DM}$) for resonant leptogenesis. Right panel : $M_{\rm in}$ vs $m_{S}$ ($m_{\rm DM}$) for baryogenesis. The semi-classical approximation is assumed to be valid till the end of PBH lifetime. The shaded region below the solid blue line is consistent with the observed baryon asymmetry. The solid magenta line denotes correct DM relic. The vertical green shaded region denotes the PBH mass window for successful cogenesis. The lower bound on DM mass from structure formation constraints is shown by the dotted magenta line. The dashed blue line corresponds to $T_{\rm ev} = m_{N_{1}}$ (left panel) and $T_{\rm ev} = m_{S}$ (right panel).}
    \label{fig:mass_scan1}
\end{figure}

\begin{figure}
    \centering
    \includegraphics[width=0.48\linewidth]{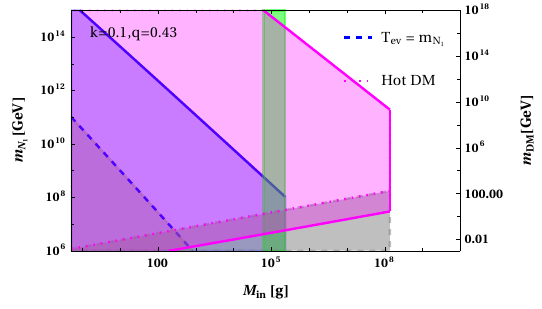}
    \includegraphics[width=0.48\linewidth]{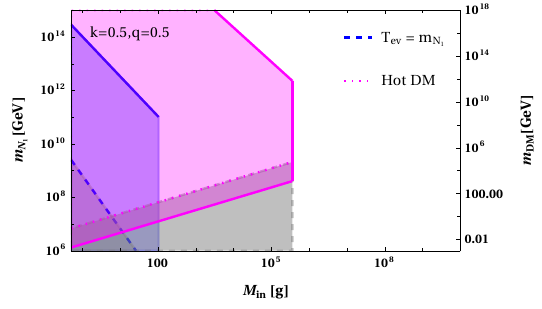}
    \includegraphics[width=0.48\linewidth]{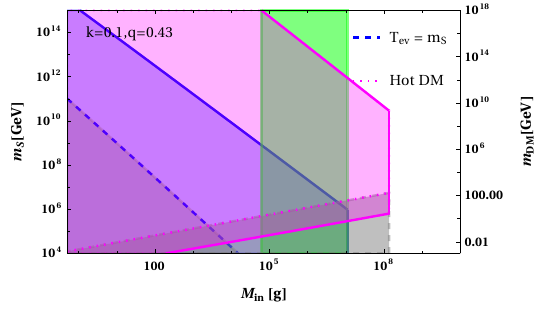}
    \includegraphics[width=0.48\linewidth]{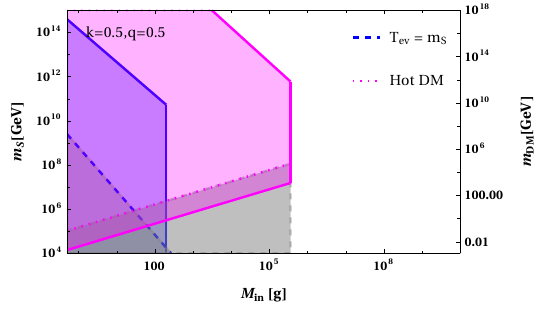}
    \caption{$M_{\rm in}$ vs $m_{N_{1}}$ ($m_{\rm DM}$) for two different set of memory-burden parameters: \{$k=0.1, q=0.43$\} and \{$k=0.5, q=0.5$\}, for resonant leptogenesis (top panels) and  baryogenesis (bottom panels). Other descriptions remain same as in Fig. \ref{fig:mass_scan1}.}
    \label{fig:mass_scan2}
\end{figure}

Fig. \ref{fig:mass_scan1} and Fig. \ref{fig:mass_scan2} summarise our cogenesis results with semi-classical and memory-burdened PBH respectively. In the semi-classical regime, cogenesis is viable both for resonant leptogenesis and baryogenesis as shown in $m_{N_1} (m_S)-M_{\rm in}-m_{\rm DM}$ parameter space on left and right panel plots respectively of Fig. \ref{fig:mass_scan1}. In both the scenarios, superheavy DM and baryon asymmetry can be produced from PBH evaporation in semi-classical regime. As expected, baryogenesis offers a wider parameter space in $M_{\rm in}$ due to the possibility of producing asymmetry as late as prior to the BBN epoch. Fig. \ref{fig:mass_scan2} shows the corresponding parameter space for two different choices of \{$k$, $q$\} controlling the departure from semi-classical approximation. While cogenesis is possible for the chosen combination of \{$k$, $q$\} in the left panels, it is not so for the right panels. 

\section{Gravitational Waves}
\label{sec3}
PBH can generate primordial gravitational waves in a variety of ways. A few notable sources are (i) the large curvature perturbations which are responsible for the formation of PBH can induce GWs \cite{Saito:2008jc}, (ii) gravitons from PBH evaporation \cite{Anantua:2008am}, (iii) PBH mergers \cite{Zagorac:2019ekv} and (iv) density fluctuations due to the inhomogeneous distribution of PBHs \cite{Papanikolaou:2020qtd, Domenech:2020ssp, Inomata:2020lmk}. Here, we consider the last possibility as the corresponding GW spectrum for ultra-light PBH remains within reach of ongoing and near future GW experiments. Also, this contribution is independent of PBH formation mechanism and free from late-time astrophysical uncertainties. After the formation of PBH, they have a random spatial distribution dictated by Poisson statistics \cite{Papanikolaou:2020qtd}. These inhomogeneities in the PBH distribution lead to density fluctuations which are isocurvature in nature. During PBH domination, the isocurvature perturbations are converted to adiabatic perturbations which further induce GW at second order. These GWs are further enhanced due to the almost instantaneous evaporation of PBH. While this has been studied in several earlier works, we follow the prescription of \cite{Papanikolaou:2020qtd, Domenech:2020ssp, Balaji:2024hpu, Barman:2024iht} to estimate the GW spectrum for memory-burdened PBH consistent with cogenesis discussed in the previous sections.

The peak amplitude of GW during PBH evaporation can be found to be
\begin{eqnarray}
    \Omega^{\rm peak}_{\rm {GW, ev}} &\simeq& \frac{1}{4133^{\frac{4}{3+2k}}}q^4\left(\frac{3+2k}{3}\right)^{-\frac{7}{3}+\frac{4}{9+6k}} \frac{\beta^{16/3}\,{\rm exp[8k(7-\frac{4}{3+2k})]}}{2.3\times10^{-20}}  \nonumber \\
     &\times& \left(\frac{q M_{\rm in}}{1 \rm g}\right)^{\frac{2}{3}(1+k)(7-\frac{4}{3+2k})} 
     \begin{cases}
          1, \text{ \hspace{0.5cm} for \hspace{0.5cm}} \beta >  \beta_{*} \hspace{0.5cm} \\ 
          q^8, \text{\hspace{0.5cm} for \hspace{0.5cm}} \beta <  \beta_{*}.
     \end{cases}
\end{eqnarray}
Here $\beta_{*}$ corresponds to that particular value of $\beta$ above which early PBH-radiation equality occurs before MB is activated. The expression of $\beta_{*}$ in terms of $M_{\rm in}$ and $q$ are given as 
\begin{eqnarray}
    \beta_{*} = \left(\frac{3 \, \epsilon}{16 \, \pi \, \gamma (1-q^3) S(M_{\rm in})}\right)^{1/2} \simeq 7.3\times10^{-6} \frac{1}{\sqrt{1-q^3}} \left(\frac{1 \, \rm g}{M_{\rm in}}\right).
\end{eqnarray}
The spectrum of GWs amplitude today can be written as
\begin{eqnarray}
    \Omega_{\rm GW,0} h^2 (f) = 1.62\times 10^{-5} \, \Omega_{\rm GW, ev}^{\rm peak} h^2 \left(\frac{f}{f_{\rm UV}}\right)^{\frac{11+10k}{3+2k}} \mathcal{I}(f,k),
\end{eqnarray}
where 
\begin{eqnarray}
    \mathcal{I}(f,k) = \int_{-\xi_{0}(f)}^{\xi_{0}(f)} ds \frac{(1-s^2)^{2}}{ (1-c^2_{s} s^{2})^{(1+\frac{2}{3+2k})}},
\end{eqnarray}
and the quantity $\xi_{0}(f)$ can be read as
\begin{eqnarray}
    \xi_{0}(f) = 
    \begin{cases}
        1, \text{\hspace{0.5 cm} for \hspace{0.5 cm}} \frac{f_{\rm UV}}{f} \geq \frac{1+c_{s}}{2 c_{s}} \\
        \frac{2 f_{\rm UV}}{f} - \frac{1}{c_{s}}, \text{\hspace{0.5 cm} for \hspace{0.5 cm}} \frac{1+c_{s}}{2 c_{s}} \geq \frac{f_{\rm UV}}{f} \geq \frac{1}{2 c_{s}} \\
        0, \text{\hspace{0.5 cm} for \hspace{0.5 cm}} \frac{1}{2 c_{s}} \geq \frac{f_{\rm UV}}{f}.
    \end{cases}
\end{eqnarray}
Here, $c_{s}$ denotes the sound speed and takes value of $\frac{1}{\sqrt{3}}$ during radiation dominated era and the frequency related to the cut-off scale is
\begin{eqnarray}
    f_{\rm UV} \simeq 4.8\times 10^{6}\, \rm{Hz} \, e^{-4k} \left(\frac{3+2k}{3}\right)^{1/6} \left(\frac{1\, \rm g}{q \, M_{\rm in}}\right)^{\frac{5}{6}+\frac{k}{3}}.
\end{eqnarray}
The GWs expressions for $\beta > \beta_{*}$ is only useful for $q>0.41$. This is because for $q<0.41$ with $\beta > \beta_{*}$, another intermediate radiation domination arise which is not considered for the derivations of GWs expressions \cite{ Balaji:2024hpu,Barman:2024iht}.

\begin{figure}
    \centering
    \includegraphics[width=0.48\linewidth]{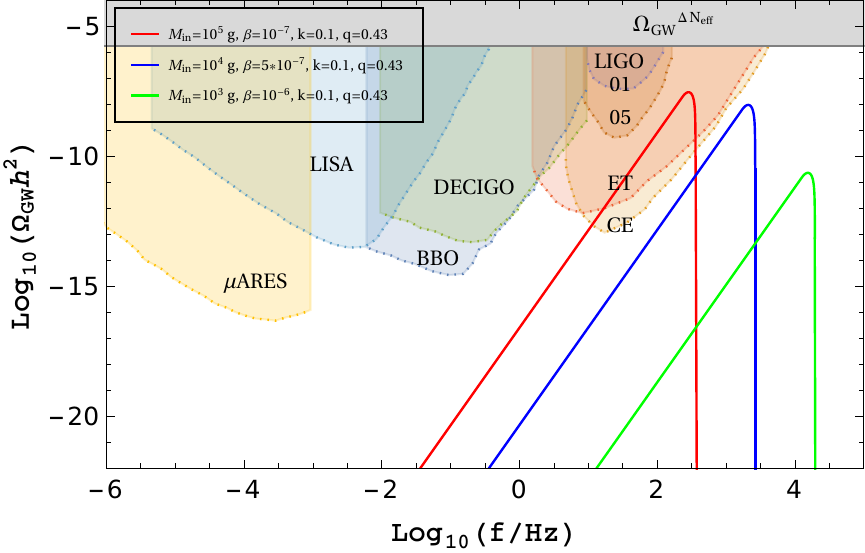}
    \includegraphics[width=0.48\linewidth]{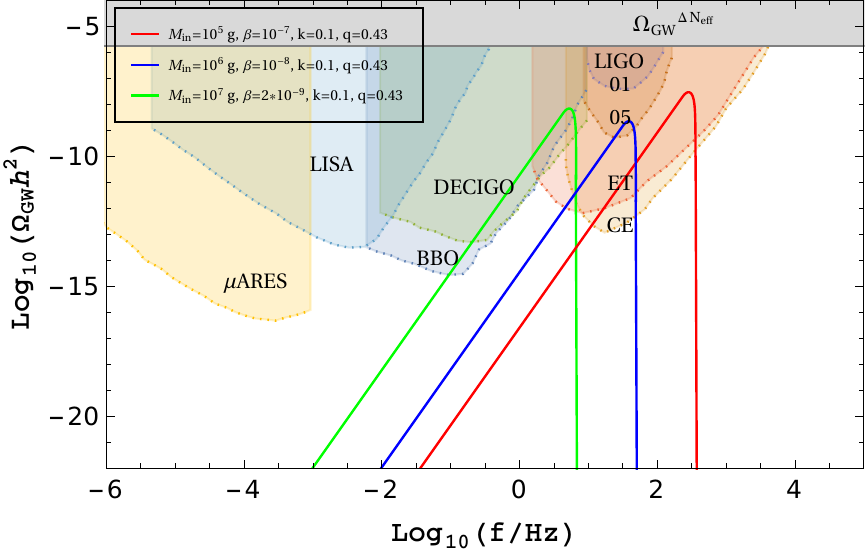}
    \caption{Gravitational wave spectrum for benchmark model parameters consistent with successful resonant leptogenesis (left panel) and baryogenesis (right panel).}
    \label{fig:GW}
\end{figure}

Fig. \ref{fig:GW} shows the GW spectrum for benchmark parameters consistent with successful resonant leptogenesis (left panel) and baryogenesis (right panel). For both the plots, we keep memory burden parameters $\{k, q\} = \{0.1, 0.43\}$ and CP asymmetry parameter $\epsilon=0.1$ fixed. While successful resonant leptogenesis requires $M_{\rm in} \leq \mathcal{O}(10^5 \, \rm g)$, baryogenesis can occur even for $M_{\rm in} \sim \mathcal{O}(10^7 \, \rm g)$. This helps in bringing the peak frequencies for all three benchmark points of successful baryogenesis within range of GW experiments. On the other hand, only one benchmark point with the largest allowed $M_{\rm in}$ in resonant leptogenesis scenario has peak frequency within range of GW experiments. The experimental sensitivities of GW detectors BBO~\cite{Crowder:2005nr,Corbin:2005ny,Harry:2006fi}, DECIGO~\cite{Seto:2001qf,Kawamura:2006up,Yagi:2011wg}, CE~\cite{LIGOScientific:2016wof,Reitze:2019iox}, ET~\cite{Punturo:2010zz, Hild:2010id,Sathyaprakash:2012jk, Maggiore:2019uih}, LISA~\cite{2017arXiv170200786A} and aLIGO/VIRGO~\cite{LIGOScientific:2014qfs, LIGOScientific:2016wof, LIGOScientific:2016jlg}, $\mu$ARES \cite{Sesana:2019vho} are shown as shaded regions of different colours. The grey shaded regions at the top of these plots are ruled out from BBN bounds on effective relativistic degrees of freedom $\Delta N_{\rm eff}$.

\begin{figure}
    \centering
    \includegraphics[width=0.45\linewidth]{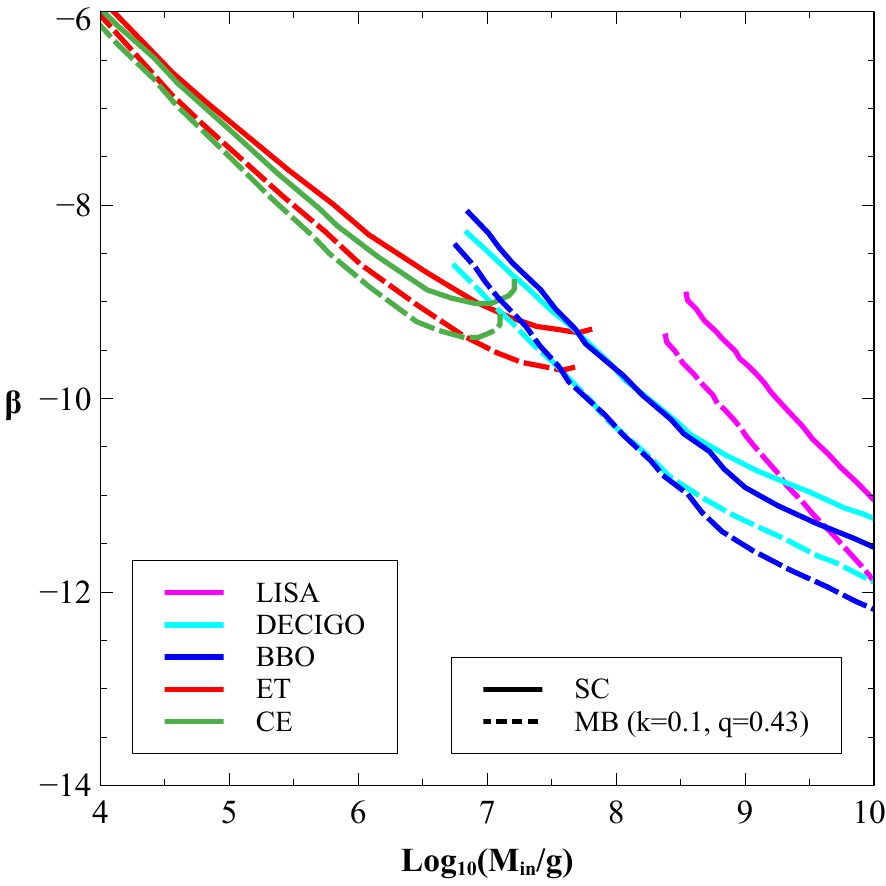}
    \caption{Translated GWs sensitivities of different future experiments on $M_{\rm in}$-$\beta$ plane. The solid lines represent the GW experiment's sensitivities for PBH obeying SC regime throughout its lifetime whereas the dashed lines represent sensitivities for memory-burdened PBH with $k=0.1$, $q=0.43$.}
    \label{fig:GW_comparison}
\end{figure}

In Fig. \ref{fig:GW}, we have shown the GW spectrum for different benchmark values along with sensitivities of future GW experiments. It is also possible to check the sensitivity of future GW detectors to the entire parameter space by focusing on the peak amplitude ($\Omega^{\rm peak}_{\rm GW}$) and cut-off frequency ($f_{\rm UV}$) instead of the full GW spectrum. While the peak amplitude is a function of PBH parameters as $\Omega^{\rm peak}_{\rm GW}=\Omega^{\rm peak}_{\rm GW}(M_{\rm in}, k, q, \beta$), the cut-off frequency $f_{\rm UV}=f_{\rm UV}(M_{\rm in}, k, q)$ does not depend upon $\beta$. Fixing the values of $k$ and $q$, it is possible to translate the GW sensitivities shown in Fig. \ref{fig:GW} into the $M_{\rm in}$-$\beta$ plane. Fig. \ref{fig:GW_comparison} shows the translated GW sensitivities of different future experiments on $M_{\rm in}$-$\beta$ plane. The sensitivities are shown for two different scenarios: (i) PBH obeying SC regime throughout its lifetime (solid lines), (ii) memory-burdened PBH  with $k=0.1$ and $q=0.43$ (dashed lines). With MB effect included, future GWs sensitivities can probe a larger portion of PBH parameter space compared to the SC regime.

\begin{figure}
    \centering
    \includegraphics[width=0.45\linewidth]{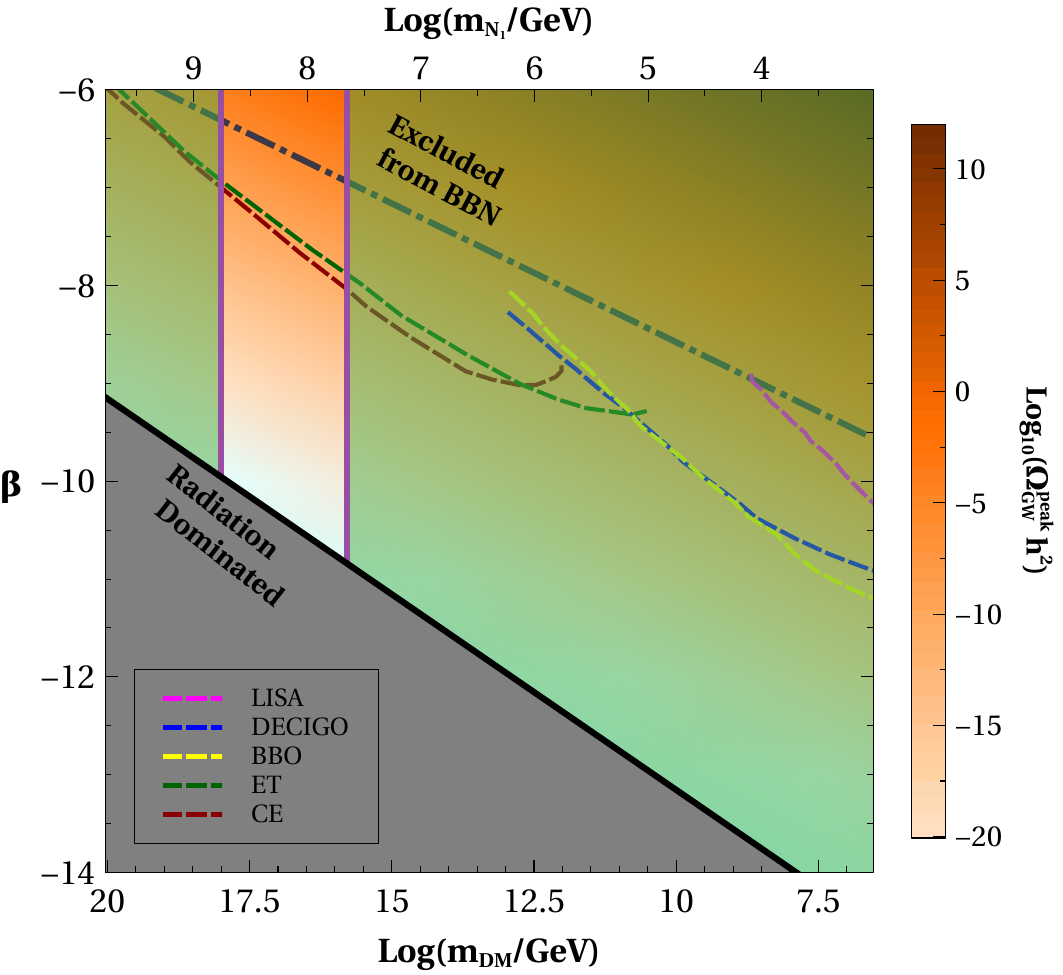}
    \includegraphics[width=0.45\linewidth]{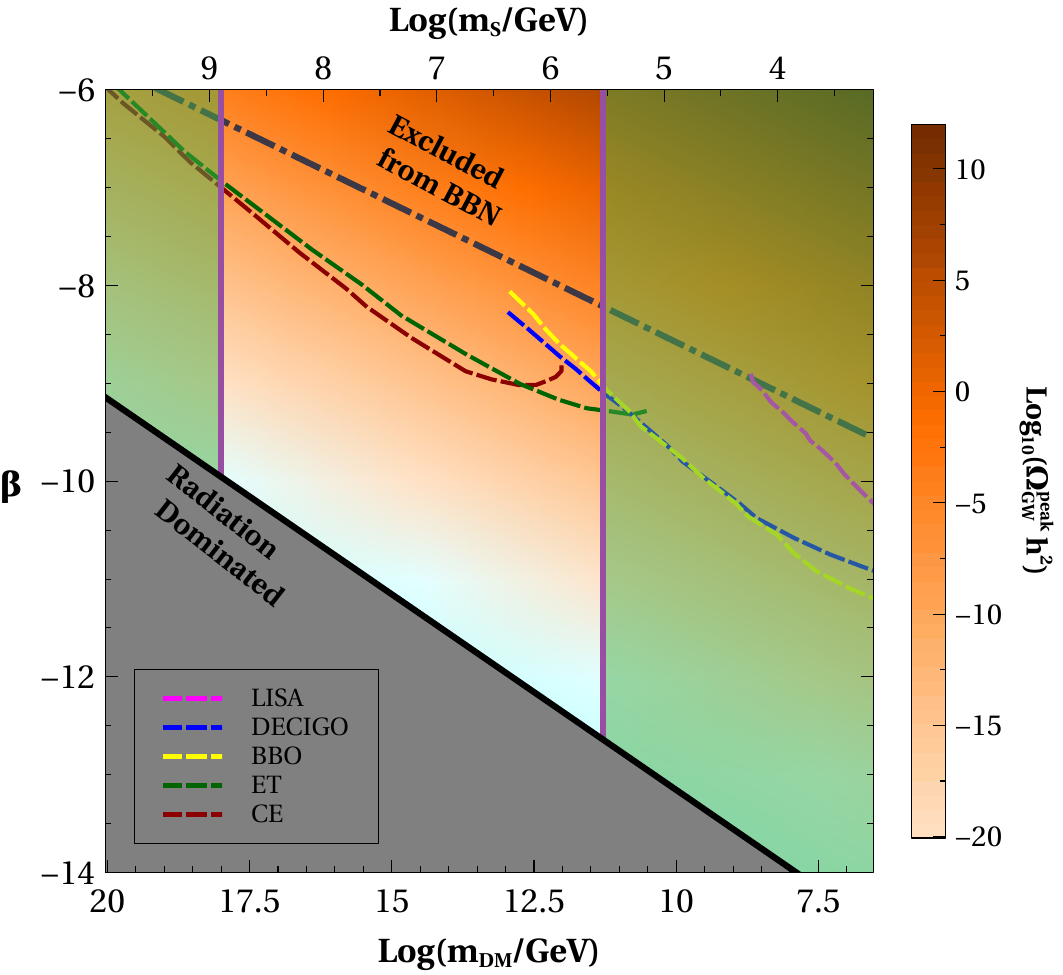}
    \caption{Parameter space plot in $m_{\rm DM} - m_{\rm N_{1}} (m_{S})-\beta$ plane for semi-classical scenario showing region that can be probed by future GWs detection experiments for cogenesis with resonant leptogenesis (left plot) and for cogenesis with bayogenesis (right plot). The vertical green shaded regions denote the parameter space where cogenesis is not possible.}
    \label{fig:GW_scan_SC}
\end{figure}

\begin{figure}
    \centering
    \includegraphics[width=0.45\linewidth]{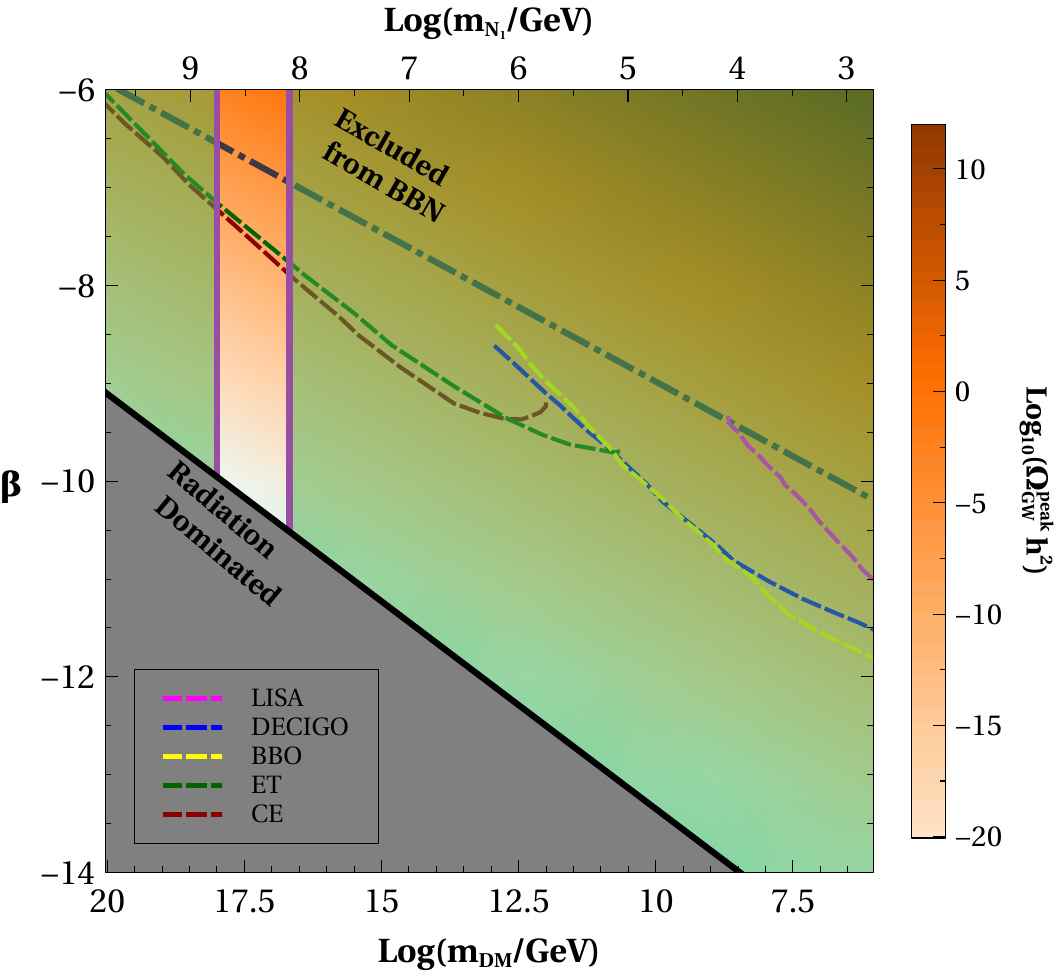}
    \includegraphics[width=0.45\linewidth]{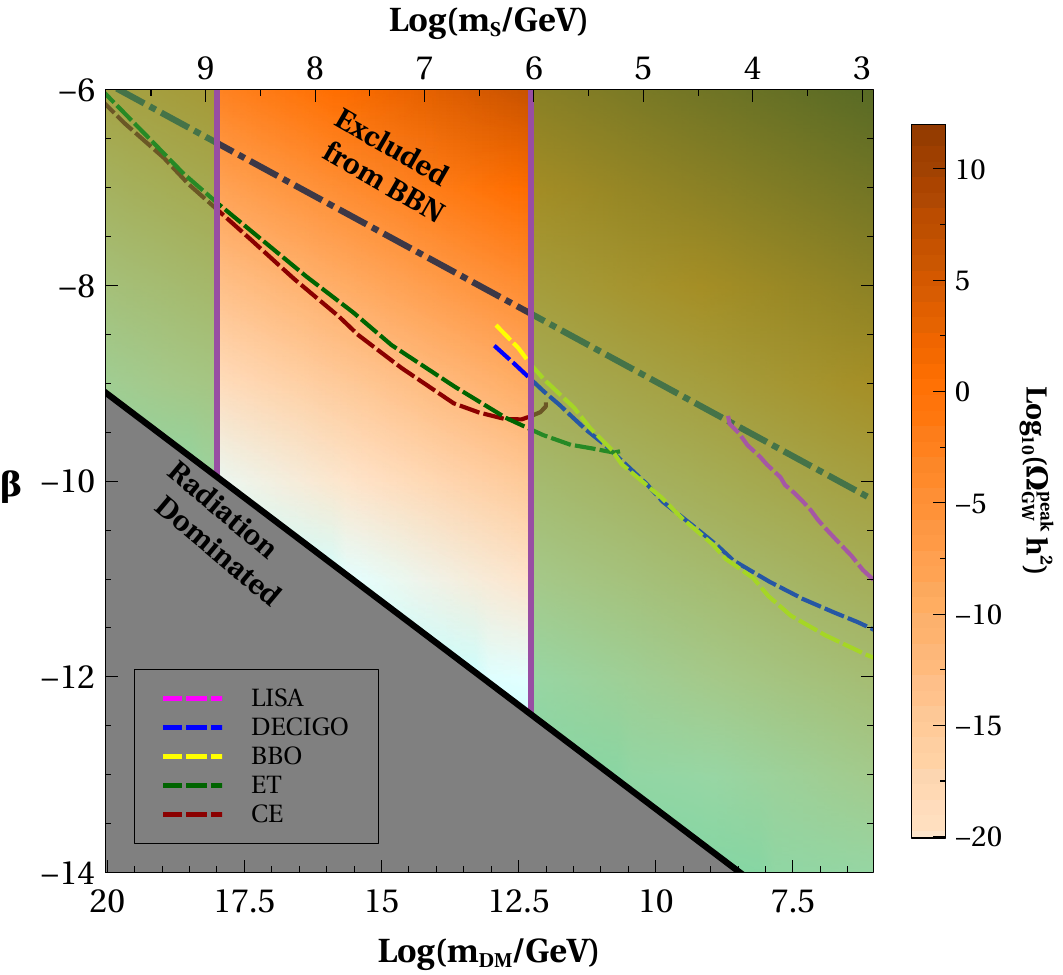}
    \caption{Parameter scan plot in $m_{\rm DM} - m_{\rm N_{1}}(m_{S})-\beta$ plane for memory-burdened scenario showing the region that can be probed by future GWs detection experiments for cogenesis with resonant leptogenesis (left plot) and for cogenesis with bayogenesis (right plot). The vertical green shaded regions denote the parameter space where cogenesis is not possible. Here we consider $k=0.1$ and $q=0.43$.}
    \label{fig:GW_scan}
\end{figure}

Further, for $\beta > \beta_{c}$, the DM abundance depends on $M_{\rm in}, m_{\rm DM}, k$ and $q$. Similarly, baryon asymmetry depends on $M_{\rm in}, m_{N_{1}} (m_{S}), k$ and $q$. Fixing $k$ and $q$, one can correlate $M_{\rm in}$ with $m_{\rm DM}$ by imposing correct DM relic condition. In the same way, $M_{\rm in}$ can be correlated with $m_{N_{1}}$ or $m_{S}$ imposing correct baryon asymmetry condition.  Fig. \ref{fig:GW_scan_SC} and $\ref{fig:GW_scan}$ show the allowed parameter space for cogenesis on $m_{\rm DM}-m_{N_{1}}-\beta$ plane for resonant leptogenesis (left) and on $m_{\rm DM}-m_{S}-\beta$ plane for baryogenesis (right). The results for SC regime and memory burdened regime (with $k=0.1$ and $q=0.43$) are shown in Fig. \ref{fig:GW_scan_SC} and Fig. \ref{fig:GW_scan} respectively. The dashed lines denote the future sensitivities whereas the upper right triangular region towards the right of the dot-dashed line is disallowed from BBN constraints. The color bar denotes the peak amplitude of GWs spectrum. As expected, cogenesis with baryogenesis opens up a much larger parameter space for future GWs detection compared to cogenesis with resonant leptogenesis. These figures also show the shrinking parameter space for cogenesis while going from the semi-classical to the memory-burdened regime of PBH.

\section{Conclusion}
\label{sec4}
We have studied the possibility of simultaneously producing baryon asymmetry and dark matter in the Universe non-thermally from evaporation of ultra-light primordial black holes by incorporating memory burden effects. Starting with one of the simplest scenario of baryogenesis via leptogenesis with hierarchical right handed neutrinos, we show that memory-burdened PBH mass required for successful leptogenesis $M_{\rm in} \leq \mathcal{O}(10 \, \rm g)$ leads to correct DM relic only for light DM, already ruled out from structure formation constraints. Considering quasi-degenerate right handed neutrinos take us to the resonant leptogenesis regime where we show that cogenesis of baryon asymmetry and cold dark matter is feasible consistent with all observational constraints. For resonant leptogenesis scenario, memory-burdened PBH mass can be as large as $M_{\rm in} \sim \mathcal{O}(10^5 \, \rm g)$ which can produce superheavy DM consistent with relic and structure formation constraints. Finally, we show the viability of directly producing baryon asymmetry together with DM which allows PBH mass as large as $M_{\rm in} \sim \mathcal{O}(10^7 \, \rm g)$. We summarise a few benchmark points for resonant leptogenesis and baryogenesis in table \ref{tab:co_lepto} and \ref{tab:co_baryo} respectively. Interestingly these two successful cogenesis scenarios can be distinguished from their gravitational wave predictions as shown in Fig. \ref{fig:GW}, \ref{fig:GW_scan_SC} and \ref{fig:GW_scan}. With several future GW experiments sensitive to the memory-burdened PBH parameter space consistent with cogenesis, the detection prospects remain very promising.

\begin{table}[!htb]
\centering
\begin{tabular}{c c c c c c c c c}
\hline \hline 
BPs & $M_{\rm in}$  (g)& $k$ & $q$ & $m_{N_{1}}$ (GeV)& $m_{\rm DM}$ (GeV) & Cogenesis \\
 \hline 
 BP1 & $10^{5}$ & $0.10$ & $0.43$ & $3\times 10^{8}$ & $3\times 10^{17}$ & Yes\\
 BP2 & $10^{4}$ & $0.10$ & $0.43$ & $6\times 10^{9}$ & $9\times 10^{-3}$ & No \\
 BP3 & $10^{3}$ & $0.10$ & $0.43$ & $1\times 10^{11}$ & $3\times 10^{-3}$ & No \\
 \hline \hline
\end{tabular}
\caption{Cogenesis with resonant leptogenesis.}
\label{tab:co_lepto}
\end{table}

\begin{table}[!htb]
\centering
\begin{tabular}{c c c c c c c c c}
\hline \hline 
BPs & $M_{\rm in}$  (g)& $k$ & $q$ & $m_{S}$ (GeV)& $m_{\rm DM}$ (GeV) & Cogenesis \\
 \hline 
 BP4 & $10^{5}$ & $0.10$ & $0.43$ & $6\times 10^{8}$ & $3\times 10^{17}$ & Yes\\
 BP5 & $10^{6}$ & $0.10$ & $0.43$ & $3\times 10^{7}$ & $7\times 10^{14}$ & Yes \\
 BP6 & $10^{7}$ & $0.10$ & $0.43$ & $2\times 10^{6}$ & $2\times 10^{12}$ & Yes \\
 \hline \hline
\end{tabular}
\caption{Cogenesis with baryogenesis}
\label{tab:co_baryo}
\end{table}

\section*{Acknowledgments}
The authors would like to thank Basabendu Barman for useful discussions. The work of DB is supported by the Science and Engineering Research Board (SERB), Government of India grants MTR/2022/000575 and CRG/2022/000603. The work of ND is supported by the Ministry of Education, Government of India via the Prime Minister's Research Fellowship (PMRF) December 2021 scheme. 

\appendix

\section{Particle production from PBH with MB}
\label{appen1}

Here, we show the derivation of Eq. \eqref{eq:Nj} which estimates the particle production from PBH incorporating memory-burden effect. To calculate the number of particles $N_{j}$ of a species $j$ with mass $m_{j}$ and intrinsic spin $g_{j}$ produced from complete evaporation of a PBH, we calculate production from both SC ($N^{\rm SC}_{j}$) and MB ($N^{\rm MB}_{j}$) regime separately. The emission rate per unit time per unit energy of particle species $j$ in the SC regime is given as 
\begin{eqnarray}
    \frac{d^2N^{\rm SC }_{j}}{dt dE} = \frac{27}{4}\frac{g_{j}}{32\pi^3} \frac{(E/T_{\rm BH})^2}{e^{E/T_{\rm BH}}\pm 1},
\end{eqnarray}
where $\pm$ signs represent fermionic and bosonic nature of species $j$ respectively. Integrating the above equation with respect to energy, we get
\begin{eqnarray}\label{eq:SC1}
    \frac{dN^{\rm SC}_{j}}{dt} = \frac{27}{4} \frac{\xi g_{j} \zeta(3)}{16 \pi^3} \frac{M^2_{\rm P}}{M_{\rm BH}}. 
\end{eqnarray}
Here $\xi$ takes the value of $3/4$ and $1$ for fermions and boson respectively. 

Similarly, for the MB regime, the production rate is given by 
\begin{eqnarray} \label{eq:MB1}
    \frac{dN^{\rm MB}_{j}}{dt} = \frac{1}{[S(M_{\rm BH})]^k}\frac{27}{4} \frac{\xi g_{j} \zeta(3)}{16 \pi^3} \frac{M^2_{\rm P}}{M_{\rm BH}} = \frac{27}{4} \frac{\xi g_{j} \zeta(3) 2^k}{16 \pi^3} \frac{M^{2+2k}_{\rm P}}{M^{1+2k}_{\rm BH}}.
\end{eqnarray}
For $m_{j}<T^{\rm in}_{\rm BH}$, production of species $j$ happens from the beginning of PBH evolution. For $m_{j}>T^{\rm in}_{\rm BH}$, production of species $j$ begins to occur only after instantaneous Hawking temperature of PBH becomes equal to $m_{j}$ at an epoch $t_{j}$, i.e. $T_{\rm BH}(t_{j})=m_{j}$. In the following, we discuss both the possibilities.

\textbf{For $\mathbf{m_{j}<T^{\rm in}_{\rm BH}}$ :} For semi-classical regime, integrating Eq. \eqref{eq:SC1} for the duration ($t_{\rm in}$,  $t_{q}$) that correspond to mass range ($M_{\rm in}, q M_{\rm in}$), we obtain 
\begin{eqnarray}\label{eq:11}
    N^{\rm SC}_{j}\Big|_{m_{j}<T^{\rm in}_{\rm BH}}=\frac{27}{128} \frac{\xi\, g_{j}\, \zeta(3)}{ \epsilon \, \pi^3} \frac{M^2_{\rm in}}{M^2_{\rm P}} (1-q^2).
\end{eqnarray}
For memory-burdened regime, we integrate Eq. \eqref{eq:MB1} during ($t_{q},t^{k}_{\rm ev}$) that correspond to PBH mass range ($qM_{\rm in}, 0$) and obtain
\begin{eqnarray}\label{eq:12}
    N^{\rm MB}_{j}\Big|_{m_{j}<T^{\rm in}_{\rm BH}} = \frac{27}{128} \frac{\xi\, g_{j}\, \zeta(3)}{ \epsilon \, \pi^3} \frac{M^2_{\rm in}}{M^2_{\rm P}} q^2.
\end{eqnarray}
Combining $N^{\rm SC}_{j}$ and $N^{\rm MB}_{j}$, the total number of particles produced becomes
\begin{eqnarray}
    N_{j}\Big|_{m_{j}<T^{\rm in}_{\rm BH}} = N^{\rm SC}_{j}\Big|_{m_{j}<T^{\rm in}_{\rm BH}} + N^{\rm MB}_{j}\Big|_{m_{j}<T^{\rm in}_{\rm BH}} = \frac{27}{128} \frac{\xi\, g_{j}\, \zeta(3)}{ \epsilon \, \pi^3} \frac{M^2_{\rm in}}{M^2_{\rm P}}.
\end{eqnarray}

\textbf{For $\mathbf{m_{j}>T^{\rm in}_{\rm BH}}$ :}  Here, the production of species $j$ starts at an epoch $t_{j}$ corresponding to the condition $T_{\rm BH}(t_{j})=m_{j}$. The production can be either in the semi-classical regime or in the memory-burdened regime. Using Eq. \eqref{eq:MBH_time_SC} and \eqref{eq:MBH_time_MB}, the expression for $t_{j}$ can be written as
\begin{eqnarray}
    t_{j} = \begin{cases}
        t_{\rm in} + \frac{1}{\Gamma^{0}_{\rm BH}}\left(1-\frac{M^6_{\rm P}}{m^3_{j}M^3_{\rm in}}\right) \,\,\,\,\,\,\,\,\,\,\,\,\,\,\, \,\,\,\,\,\,\,\,\,\,\,\,\text{for } t_{j}<t_{q},\\
        t_{\rm q} + \frac{1}{\Gamma^{k}_{\rm BH}}\left(1-\left(\frac{M^2_{\rm P}}{q\,m_{j}M_{\rm in}}\right)^{3+2k}\right) \,\,\,\,\,\,\, \text{for } t_{j}>t_{q}.
    \end{cases}
\end{eqnarray}
When $t_{j}<t_{q}$, there is no production of species $j$ in the period ($t_{\rm in}$,$t_{j}$). From $t_{j}$ to $t_{q}$, integrating Eq. \eqref{eq:MBH_time_SC}, the number of produced particles is found as 
\begin{eqnarray}
    N^{\rm SC}_{j}\Big|^{t_{j}<t_{q}}_{m_{j}>T^{\rm in}_{\rm BH}} = \frac{27}{128} \frac{\xi\, g_{j}\, \zeta(3)}{ \epsilon \, \pi^3} \frac{M^2_{\rm in}}{M^2_{\rm P}} \left(\left(\frac{M^2_{\rm P}}{m_{j}M_{\rm in}}\right)^2-q^2\right).
\end{eqnarray}
The number $N^{\rm MB}_{j}$ remains same as in \eqref{eq:12} and is given by 
\begin{eqnarray}
    N^{\rm MB}_{j}\Big|^{t_{j}<t_{q}}_{m_{j}>T^{\rm in}_{\rm BH}} = \frac{27}{128} \frac{\xi\, g_{j}\, \zeta(3)}{ \epsilon \, \pi^3} \frac{M^2_{\rm in}}{M^2_{\rm P}} q^2,
\end{eqnarray}
as species $j$ is produced throughout the MB regime. Hence, for $t_{j}<t_{q}$,
\begin{eqnarray}
    N_{j}\Big|^{t_{j}<t_{q}}_{m_{j}>T^{\rm in}_{\rm BH}} = N^{\rm SC}_{j}\Big|^{t_{j}<t_{q}}_{m_{j}>T^{\rm in}_{\rm BH}} + N^{\rm MB}_{j}\Big|^{t_{j}<t_{q}}_{m_{j}>T^{\rm in}_{\rm BH}} = \frac{27}{128} \frac{\xi\, g_{j}\, \zeta(3)}{ \epsilon \, \pi^3} \frac{M^2_{\rm P}}{m^2_{j}}.
\end{eqnarray}
For $t_{j}>t_{q}$, there is no production of species $j$ in the semi-classical regime. For memory-burdened regime, integrating Eq. \eqref{eq:MBH_time_MB} from $t_{j}$ to $t^{k}_{\rm ev}$, we obtain 
\begin{eqnarray}
     N^{\rm MB}_{j}\Big|^{t_{j}>t_{q}}_{m_{j}>T^{\rm in}_{\rm BH}} = \frac{27}{128} \frac{\xi\, g_{j}\, \zeta(3)}{ \epsilon \, \pi^3} \frac{M^2_{\rm P}}{m^2_{j}}.
\end{eqnarray}
This further gives
\begin{eqnarray}
    N_{j}\Big|^{t_{j}>t_{q}}_{m_{j}>T^{\rm in}_{\rm BH}} = N^{\rm MB}_{j}\Big|^{t_{j}>t_{q}}_{m_{j}>T^{\rm in}_{\rm BH}} = \frac{27}{128} \frac{\xi\, g_{j}\, \zeta(3)}{ \epsilon \, \pi^3} \frac{M^2_{\rm P}}{m^2_{j}}.
\end{eqnarray}
Note that for both $t_{j}<t_{q}$ and $t_{j}>t_{q}$, the final produced amount of species $j$ remains the same i.e. $ N_{j}\Big|^{t_{j}<t_{q}}_{m_{j}>T^{\rm in}_{\rm BH}} = N_{j}\Big|^{t_{j}>t_{q}}_{m_{j}>T^{\rm in}_{\rm BH}} \equiv N_{j}\Big|_{m_{j}>T^{\rm in}_{\rm BH}}$.

\section{DM and baryon asymmetry from gravity mediated process}
\label{appen3}

Apart from production due to PBH evaporation, there is an unavoidable production channel of particles from 2-to-2 annihilations of SM particles via the exchange of massless gravitons. The Boltzmann equation of a  particle with mass $m_{j}$ produced via gravitational UV freeze-in is given by \cite{Bernal:2018qlk}
\begin{eqnarray}
    \frac{d n_{j}}{dt} + 3 \mathcal{H} n_{j} = \delta \frac{T^{8}}{M^4_{\rm P}},
\end{eqnarray}
where $\delta$ takes the value of $1.9\times 10^{-4}$ for real scalar and $1.1\times 10^{-3}$ for Dirac fermion. For $m_{j}<T_{\rm RH}$ (with $T_{\rm RH}$ being the reheat temperature) the freeze-in comoving abundance of species $j$ can be obtained by integrating the above equation from reheat temperature $T_{\rm RH}$ to a much lower temperature  $T$. If $m_{j}$ is larger than reheat temperature (but smaller than the maximum temperature reached by the SM bath), the production occurs during the reheating period itself. For both the scenarios, the freeze-in comoving abundance of species $j$ is given as
\begin{eqnarray}
    Y_{0\,j} = \frac{45 \times \delta}{2 \,\pi^3 \, g_{*\,s}}\sqrt{\frac{10}{g_{*}}}
    \begin{cases}
         \frac{T^3_{\rm RH}}{M^3_{\rm P}} \,\,\,\,\,\,\,\, \text{for $m_{j}<T_{\rm RH}$},\\
         \frac{T^7_{\rm RH}}{M^3_{\rm P} m^4_{j}} \,\,\,\,\,\,\,\, \text{for $m_{j}>T_{\rm RH}$}.
    \end{cases}
\end{eqnarray}
The produced amount of particles will get diluted due to PBH evaporation which can be approximated (for $\beta > \beta_{c}$) as
\begin{eqnarray}
    \frac{S(T_{\rm in})}{S(T_{\rm ev})} \approx \frac{T_{\rm ev}}{T^{\rm BH}_{\rm eq}},
\end{eqnarray}
where $T^{\rm BH}_{\rm eq}$ is defined when PBH energy density becomes equal to radiation energy density after PBH formation $\rho_{\rm rad}(T^{\rm BH}_{\rm eq})=\rho_{\rm BH}(T^{\rm BH}_{\rm eq})$ and can be expressed as $T^{\rm BH}_{\rm eq}\simeq \beta \, T_{\rm in}$. With these, the DM abundance from such gravity mediated process can be given as 
\begin{eqnarray}
    \Omega_{\rm DM} h^2 = 2.56\times 10^{8}\, \frac{m_{\rm DM}}{\rm GeV} \,Y_{0\, \rm DM} \,  \frac{S(T_{\rm in})}{S(T_{\rm ev})}.
\end{eqnarray}
Similar to DM production, gravity mediated process can give rise baryon asymmetry both via leptogenesis as well as baryogenesis. For leptogenesis, its contribution can be quantified as
\begin{eqnarray}
    Y^{\Delta L}_{\rm B} (T_{0}) = \epsilon^{\Delta L}_{1}\, a_{\rm sph} Y_{0\, \rm N_{1}} \, \frac{S(T_{\rm in})}{S(T_{\rm ev})}, 
\end{eqnarray}
where $Y_{0\, N_{1}}$ describes UV freeze-in comoving abundance of $N_{1}$ from gravity mediated process. Similarly, its contribution to baryogenesis can be given as
\begin{eqnarray}
   Y_B(T_0)  = (\epsilon_{1} + \epsilon_{2}) Y_{0\, \rm S} \, \frac{S(T_{\rm in})}{S(T_{\rm ev})},
\end{eqnarray}
where $Y_{0\, S} \equiv Y_{0\, S_{1}}=Y_{0\, S_{2}}$ describes UV freeze-in comoving abundance of $S_{1,2}$ from gravity mediated process.

\begin{figure}[htb]
    \centering
    \includegraphics[width=0.48\linewidth]{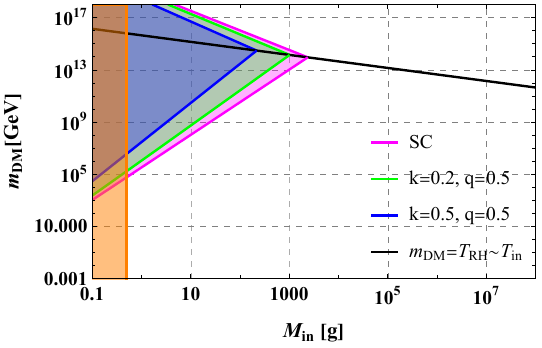}
    \includegraphics[width=0.48\linewidth]{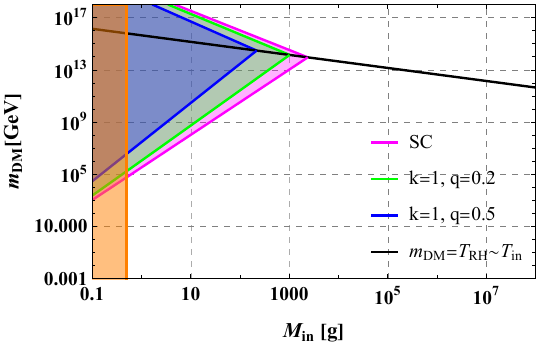}
    \caption{$m_{\rm DM}$-$M_{\rm in}$ plot considering only gravity mediated process for different values of $\{k,q\}$. Here we consider $T_{\rm RH}\sim T_{\rm in}$ and $\beta=10^{-4}$. The black solid line represents $m_{\rm DM}=T_{\rm RH}$ and the vertical orange region denotes the PBH mass that is inconsistent with CMB data.}
    \label{fig:GM_DM}
\end{figure}

Fig. \ref{fig:GM_DM} shows the DM mass as a function of PBH mass considering gravity mediated production only. Correct DM abundance is satisfied at the boundaries of the shaded regions. Here we consider the reheating temperature to be of same order as the formation temperature of PBH: $T_{\rm RH} \sim T_{\rm in}$. The black solid lines in both the panels denote $m_{\rm DM}=T_{\rm RH}$. In both the panels, $\beta$ is kept at $10^{-4}$. As expected, the regions for memory burdened PBH shrinks compared to the semi-classical one due to larger entropy dilution. 

Similar to Fig. \ref{fig:GM_DM}, results of leptogenesis and baryogenesis can be shown for gravity mediated process. However, we find that successful leptogenesis and baryogenesis require $M_{\rm in} \lesssim 10^{-2}$ g, which is already excluded from CMB data. Hence, we have not shown the counterpart of Fig. \ref{fig:GM_DM} for leptogenesis and baryogenesis.

\section{Details of structure formation constraints}
\label{appen2}
The average kinetic energy of DM at the epoch of matter-radiation equality can be written as 
\begin{eqnarray} \label{eq:FSL_1}
    \langle E_{\rm DM}(t_{\rm eq})\rangle = \langle E_{\rm DM}(t_{\rm ev})\rangle \frac{a_{\rm ev}}{a_{\rm eq}} = \langle E_{\rm DM}(t_{\rm ev})\rangle \frac{T_{\rm eq}}{T_{\rm ev}} \left(\frac{g_{*\,s}(T_{\rm eq})}{g_{*\,s}(T_{\rm ev})}\right)^{1/3}. 
\end{eqnarray}
The average kinetic energy associated with DM at PBH evaporation can be approximately estimated as 
\begin{eqnarray}
    \langle E_{\rm DM}(t_{\rm ev})\rangle \simeq  \delta \, T_{\rm BH}(q M_{\rm in}).
\end{eqnarray}
We take $\delta$ to be 1.3 \cite{Lennon:2017tqq}. Hence, Eq. \eqref{eq:FSL_1} becomes
\begin{eqnarray} \label{eq:FSL2}
    \langle E_{\rm DM}(t_{\rm eq})\rangle \approx 2\times10^{-9} \text{GeV} \frac{M_{\rm P}}{q M_{\rm in}} \frac{\sqrt{3\times q^{3+2k}+(1-q^3)2^{k}(3+2k)\left(\frac{M_{P}}{M_{\rm in}}\right)^{2k}}}{\sqrt{3\times2^{k}(3+2k)\epsilon \left(\frac{M_{P}}{M_{\rm in}}\right)^{3+2k}}} \left(\frac{g_{*\,s}(T_{\rm eq})}{g_{*\,s}(T_{\rm ev})}\right)^{1/3}.
\end{eqnarray}
Now, the average kinetic energy of DM at matter-radiation equality provides a bound on DM mass. This can be obtained by comparing the current DM velocity with the translated lower bound on the present velocity of a thermal warm dark matter candidate. The DM velocity at present epoch can be written as
\begin{eqnarray}\label{eq:dm0}
    v_{\rm DM, 0} = \frac{p_{0}}{m_{\rm DM}} = \frac{p_{\rm eq}}{m_{\rm DM}}\frac{a_{\rm eq}}{a_{0}} \approx \frac{a_{\rm eq}}{a_{0}} \frac{\langle E_{\rm DM}(t_{\rm eq})\rangle}{m_{\rm DM}} \approx 1.8\times 10^{-4} \frac{\langle E_{\rm DM}(t_{\rm eq})\rangle}{m_{\rm DM}},
\end{eqnarray}
where we use $\frac{a_{\rm eq}}{a_{0}} \simeq 1.8\times 10^{-4}$. On the other hand, lower bound on the mass of a thermal warn DM candidate is $3.5$ keV \cite{Irsic:2017ixq}. Hence the translated lower bound on the present velocity of thermal warm DM is
\begin{eqnarray}
    v_{\rm WDM} = \frac{a_{\rm dec}}{a_{0}} \frac{\langle E_{\rm DM}(t_{\rm dec})\rangle}{m_{\rm WDM}}.
\end{eqnarray}
We consider the decoupling temperature as $T_{\rm dec} \sim m_{\rm WDM}$ which gives $\langle E_{\rm DM}(t_{\rm dec})\rangle \sim T_{\rm dec} \sim m_{\rm WDM}$. Hence, the lower bound on warm dark matter mass is 
\begin{eqnarray} \label{eq:wdm}
    v_{\rm WDM} \approx \frac{a_{\rm dec}}{a_{0}} \lesssim 1.8 \times 10^{-8}.
\end{eqnarray}
Comparing Eq. \eqref{eq:dm0} with Eq. \eqref{eq:wdm}, we get 
\begin{eqnarray}
    \langle E_{\rm DM}(t_{\rm eq})\rangle \lesssim 10^{-4} \, m_{\rm DM}.
\end{eqnarray}
Finally, using the above inequalities in Eq. \eqref{eq:FSL2}, the constraint on DM mass from structure formation can be computed as
\begin{eqnarray}
   \frac{m_{\rm DM}}{1 \,\text{GeV}} \gtrsim 2\times10^{-5} \frac{M_{\rm P}}{q M_{\rm in}} \frac{\sqrt{3\times q^{3+2k}+(1-q^3)2^{k}(3+2k)\left(\frac{M_{P}}{M_{\rm in}}\right)^{2k}}}{\sqrt{3\times2^{k}(3+2k)\epsilon \left(\frac{M_{P}}{M_{\rm in}}\right)^{3+2k}}} \left(\frac{g_{*\,s}(T_{\rm eq})}{g_{*\,s}(T_{\rm ev})}\right)^{1/3}.
\end{eqnarray}

\bibliographystyle{JHEP}
\bibliography{ref, ref1, ref2, ref3}

\end{document}